\documentclass{article}

\PassOptionsToPackage{numbers, compress}{natbib}
\usepackage[preprint]{neurips_2024}

\usepackage{natbib}
\usepackage[utf8]{inputenc} 
\usepackage[T1]{fontenc}    
\usepackage{hyperref}       
\usepackage{url}            
\usepackage{booktabs}       
\usepackage{amsfonts}       
\usepackage{nicefrac}       
\usepackage{microtype}      
\usepackage{xcolor}         
\usepackage{algorithm}
\usepackage{algorithmic}

\usepackage{amsmath,amssymb}   
\usepackage{mathtools}
\usepackage{amsthm}
\usepackage{subcaption}

\usepackage{hyperref}
\newcommand{\acro}[1]{\textsc{\MakeLowercase{#1}}}

\newcommand{\bftab}{\fontseries{b}\selectfont}
\usepackage{url}

\title{A Dynamic, Ordinal Gaussian Process Item Response Theoretic Model}

\author{%
  Yehu Chen, Jacob Montgomery, Roman Garnett \\
  Washington University in St Louis\\
\texttt{chenyehu,jacob.montgomery,garnett@wustl.edu}
}

\begin{document}

\maketitle

\begin{abstract}
Social scientists are often interested in using ordinal indicators to estimate latent traits that change over time. Frequently, this is done with item response theoretic (\acro{IRT}) models that describe the relationship between those latent traits and observed indicators. We combine recent advances in Bayesian nonparametric \acro{IRT}, which makes minimal assumptions on shapes of item response functions, and Gaussian process time series methods to capture dynamic structures in latent traits from longitudinal observations. We propose a generalized dynamic Gaussian process item response theory (\acro{GD-GPIRT}) as well as a Markov chain Monte Carlo sampling algorithm for estimation of both latent traits and response functions. We evaluate \acro{GD-GPIRT} in simulation studies against baselines in dynamic \acro{IRT}, and apply it to various substantive studies, including assessing public opinions on economy environment and congressional ideology related to abortion debate.
\end{abstract}

\section{Introduction}
How do the issue positions of the Congress evolve over time? Is there growing dissatisfaction with the economy after recessions? Are patients emotionally stable after psychological therapies? Answering these questions requires dynamic measures of traits or attitudes. Since self-reported ratings are known to be sensitive to individual variances and inconsistency \cite{wilcox1989some}, social scientists often rely on latent variable models to answer these questions, where the latent variable of interest is inferred from a collection of noisy categorical indicators such as survey responses, voting outcomes or event counts.

However, analyzing how these traits change over time in practice introduces two problems.  First, researchers must ensure that the inferred latent traits are comparable over time. It is widely understood that failure to do so can result in misleading or even nonsensical inferences \cite{bollen1980issues}, and this problem is particularly difficult in the presence of repeated observations of the same items. Second, scholars must make model assumptions about the functional relationship between the latent traits and observed indicators. In practice, the assumed functional forms tend to be fairly restrictive (e.g., generalized linear models), which can lead to estimation that is  biased or inefficient for real data.

In this work, we propose a generalized dynamic Gaussian process item response theory (\acro{GD-GPIRT}) for longitudinal and ordinal observations. While item response theory has seen applications in machine learning area such as predicting user preference in recommendation systems \cite{chen2005personalized, baylari2009design}, students' answer in educational testing \cite{bergner2012model, cheng2019dirt, park2023comparing} and evaluating  different machine learning methods \cite{lalor2016building, martinez2019item}, we focus on estimation of latent traits. We exploit recent advances in Bayesian nonparametric \acro{IRT} with Gaussian process priors for flexibly modeling the response functions, and Gaussian process time series to capture dynamic structures in latent traits while maintaining measurement comparability. We also propose an efficient Markov chain Monte Carlo sampling algorithm, whose effectiveness is demonstrated in simulation studies. Finally, we apply \acro{GD-GPIRT} to substantive studies: assessing public opinions on economy environment and congressional ideology leaning on abortion debate.

Our model makes two contributions to the field. First, it extends recent advances in Bayesian nonparametric models of latent traits to ordered categorical indicators. The existing models in this family are limited to continuous \cite{lawrence2003gaussian} or binary \cite{duck2020gpirt} indicators. However, in fields such as psychology or survey research, ordered-categorical responses are much more common. Other standard tools for ordered categorical responses assume strict parametric functional forms for the item response functions (\acro{IRF}s) \cite{roberts2000general, duck2022ends}, strict monotonicity \cite{molenaar1997nonparametric,van2007mokken}, or both \cite{mokken2011theory}. In contrast, \acro{GD-GPIRT} offers a compromise, allowing flexibility in specification of prior structures to control the shapes (e.g., non-monotonic, asymmetric) of \acro{IRF}s .  

Second, \acro{GD-GPIRT} provides a natural and comparable way to encode dynamics in the latent traits. Instead of estimating latent variables at different time periods independently, \acro{GD-GPIRT} models the trajectory of each unit jointly through time using latent space models with dynamic structures and controls the bandwidth of variation in time and latent space using kernel parameters. Some existing \acro{IRT}-like models also allow latent traits to move over time, by assuming these trends are low-order polynomials \cite{poole2001dwnominate, bailey2007comparable, proust2022modeling} or realizations of a Wiener process \cite{martin2002dynamic, wang2013bayesian, schnakenberg_fariss_2014, chung2015recurrent}. While the former can be far too restrictive, the latter may face a variance explosion issue in prediction due to their non-stationary nature. For example, the well-known Martin--Quinn scores for Supreme Court ideology can lead to extreme scores for justices at the ending of their careers due to this unbalanced model structure \cite{martin2002dynamic}. Alternatively, \acro{GD-GPIRT} encodes the reasonable expectation that the latent trends are centered and stationary \emph{a priori}, reducing the risk of poor identification due to scaling variance \cite{jackman2001multidimensional}. As shown in our experiments, \acro{GD-GPIRT} estimates are superior in terms of model fit and measurement quality.

\section{Related work}

\paragraph{Item response theory (\acro{IRT}).} \acro{IRT} is a popular latent variable framework in social science for computerized adaptive testing \cite{xu2006computerized}, survey experiments \cite{muraki1990fitting, olino2012depression}, and political ideology scaling \cite{poole2000congress,poole2001dwnominate,bafumi2005practical}. Classic static and binary \acro{IRT} methods usually make parametric assumptions on the shapes of \acro{IRF}s, such as logistic relation (\acro{2PL} and \acro{3PL}) for monotonic and symmetric \acro{IRF}s \cite{molenaar1997nonparametric,mokken2011theory}, graded unfolding structure for non-monotonicity \cite{roberts2000general}, and logistic positive exponential family for asymmetry \cite{samejima2000logistic}. Non-parametric \acro{IRT} such as Mokken scaling \cite{mokken2011theory}, monotone unidimensional model \cite{holland1986conditional}, and dimensionality assessment model \cite{stout1987nonparametric} has also emerged to address potential mis-modeling of parametric \acro{IRT} \cite{junker2001nonparametric}. Previous work on machine learning for \acro{IRT} \cite{chen2019beta,cheng2019dirt, nguyen2022spectral} largely target prediction of unseen responses of static \acro{IRT} and have not fully explored the area of latent variable measurement for dynamic \acro{IRT}. Lastly, existing \acro{IRT} models for ordered-categorical responses are mostly parametric, including graded response model \cite{samejima1997graded}, graded unfolding models \cite{roberts1996graded, roberts2000general}, generalized partial credit model \cite{muraki1992generalized} and more \cite{agresti1990categorical,zumbo2007ordinal,van2011ordinal,bacci2014class, asparouhov2018dynamic, mcneish2023dynamic}, while leaving non-parametric \acro{IRT} under-explored. 


\paragraph{Dynamic latent variable models.}  \acro{IRT} has been widely applied as a latent variable method to learn temporal shifts in latent traits. In political science, \citet{poole1985spatial} proposed  an ideal-point spatial model (\acro{NOMINATE}) for scaling congressional roll-call votes. Their model has been extensively used in studies of Congress, and was later extended to analyze ideological trends over multiple sessions by either assuming a simple polynomial time series model \cite{poole2000congress} or estimating each session separately \cite{nokken2004congressional}. In legal studies, \citet{martin2002dynamic} proposed a dynamic Bayesian measurement model (\acro{D-IRT}) based on Bayesian random walk priors to study case dispositions of the U.S.\ Supreme Court, which was extended to ordinal responses by \citet{schnakenberg_fariss_2014} to study governmental respects of human rights. In educational research, \citet{wang2013bayesian} applied dynamic linear models to \acro{IRT} for computerized adaptive testing. There are other latent variable models for analyzing longitudinal panel data, such as the growth curve model \cite{rogosa1982growth, curran2003use, masyn2014growth} and autoregressive latent trajectory model \cite{bollen2004autoregressive, hamaker2005conditions, bollen2010overview}, but these methods limit their modeling of latent traits to either strict linearity \cite{rogosa1982growth, poole2001dwnominate, hamaker2005conditions, bollen2010overview, wang2013bayesian} or non-smooth autoregression \cite{martin2002dynamic, bollen2004autoregressive, hamaker2005conditions, bollen2010overview, schnakenberg_fariss_2014} with strong assumptions.

\paragraph{Gaussian Process Latent Variable Model (\acro{GPLVM}).} \acro{GPLVM} is a family of algorithms that reduce high-dimensional data into low-dimensional embeddings \cite{lawrence2003gaussian, lawrence2005probabilistic} and has been utilized for data visualization \cite{jiang2012supervised} and manifold learning \cite{urtasun2007discriminative, titsias2010bayesian, gao2010supervised}. Relatedly, {Gaussian Process Dynamical Systems} (\acro{GPDM}) \cite{damianou2011variational} generalizes  \acro{GPLVM} to time-series data by incorporating a temporal latent variable layer and a hidden functional layer \cite{wang2005gaussian, lawrence2007hierarchical, damianou2011variational}. However, directly applying either \acro{GPLVM} or \acro{GPDM} to \acro{IRT} is not appropriate because they typically marginalize out either the mappings or the latent variables \cite{lawrence2005probabilistic, damianou2011variational} and optimize the other, while \acro{IRT} requires inference of both. Previous attempts of integrating \acro{GPLVM} for static \acro{IRT} include \citet{urtasun2007discriminative} and \citet{duck2020gpirt} that introduced a Bayesian non-parametric Gaussian process \acro{IRT} to analyze voting behaviors of the U.S.\ Congress, but significant research gap still persists in exploiting \acro{GPLVM} for dynamic \acro{IRT} with ordinal responses. 



 \section{Problem setup}
We first setup the problem of ordinal item response theory, and then discuss existing approaches in the dynamic setting with time-series data. 

\subsection{Ordinal item response theory}
Consider the case of $n$ respondents answering $m$ different items, where response $y_{ij}$ ($i=1,\dots,n,j=1,\dots,m$) of the $i$th respondent and $j$th item belongs to an ordered category set $\mathcal{Y}_j=\{1,2,\dots,C_j\}$ with total $C_j$ levels. For example, all $C_j$ will be 5 in the five-level Likert scale \cite{likert1932technique}, where respondents may choose from ``strongly (dis)agree'', ``(dis)agree'' and ``neutral''. Item response theory states that the likelihood of observing $y_{ij}$ is jointly determined by some respondent-level latent trait or ability score $x_{i}\in \mathcal{X}=\mathbb{R}$
and item-level response function (\acro{IRF}) $f_{j}:\mathcal{X} \rightarrow \mathcal{Y}_j$. Dropping subscripts momentarily, the likelihood of observing level $c$ is modeled as an ordered logistic with discrimination and difficulty parameter ${\beta}_{1}$ and ${\beta}_{0}$:
\begin{gather}
    \boldsymbol{\beta} = [{\beta}_{0},{\beta}_{1}]^T; \hspace{.5cm}
     f{(x;\boldsymbol{\beta})} = {\beta}_{1} x + {\beta}_{0}\\
\label{likelihood}
     p\big(y=c\mid f, \{b_c\} \big)=\Phi(b_{c-1}-f)-\Phi(b_{c}-f)
\end{gather}
where $\Phi(z)$ represents the cumulative density function of a standard normal. In addition, the latent function space is subset into $C$ intervals, whose end points are denoted by $C+1$ ordered threshold parameters $b_0<b_1<\dots<b_C$. The interval $(b_{c-1},b_{c}]$ on which the value of the function falls represents the range for the $c$th category. While $b_0=-\infty$ and $b_C=\infty$ are fixed, $b_1$ to $b_{C-1}$ can move freely under the ordered constraint. Intuitively, these $\{b_{c}\}$s control the shape of the categorical likelihood given latent function value. Note that all $\boldsymbol{\beta}$s and $\{b_{c}\}$s can be further indexed by $j$ to represent their dependency on items, and determined by maximizing the joint likelihood $\prod_{i}\prod_{j} p\big(y_{ij} \mid f_{j}(x_{i};\boldsymbol{\beta}_j), \{b_{jc}\}\big)$.

Fundamentally, item response theory contrasts with supervised learning in that the latent variables as inputs to the unknown functional mappings also need to be learned. For identification of parameters in \acro{IRT}, additional constraints are usually imposed to tackle scale variance \cite{jackman2001multidimensional}: non-Bayesian approaches either constrain the latent variables to some hypercube or set specific latent variables to fixed ``anchoring'' values, whereas Bayesian approaches typically put a constraining prior over the latent variables. We will rely on the latter for identification (see Sec. \ref{sec:gdgpirt}).

\subsection{Dynamic item response theory}
\label{subsec:dynamic_irt}

In dynamic \acro{IRT}, respondents repeatedly answer potentially different sets of questions over multiple time periods, while their latent traits may evolve over time. For exposition, we assume items are different across time, as static items are special cases. We append a new index for time $t$ to the latent trait $x_{it}$ and \acro{IRF} $f_{jt}$, as well as its parameters $\boldsymbol{\beta}_{jt}$, while unit and item are still indexed by $i$ and $j$.

Some prior work simply estimated $x_{it}$ separately for each time period, in the case of ideology estimation in Congress, known as \acro{Nokken--Poole} scores. Subsequent work proposed a polynomial model for the dynamic latent traits $x_{it}\sim \text{poly}(t)$ although it turns out that a degree of $1$ is sufficient for capturing the majority of changes in the dynamic latent positions \cite{poole2001dwnominate, bailey2007comparable}. Bayesian non-parametric methods were also introduced by utilizing the autoregressive (\acro{AR}) model or Wiener process \cite{martin2002dynamic, wang2013bayesian, schnakenberg_fariss_2014}. Broadly speaking, these non-parametric methods simplifies the inference of the entire trajectory to that at a single time point based on an informative prior:
\begin{gather}
    x_{i,1} \sim \mathcal{N}(0, C_{i}),\qquad  x_{i,t} \sim \mathcal{N}(x_{i,t-1},\sigma^2_t), \quad \forall \quad t=2,\dots,T
    \label{eq:autoregression}
\end{gather}
where $\mathcal{N}(0, C_{i})$ is the anchoring prior for the dynamic trait trajectory at time $t=1$ and $\sigma^2_t$ is the diffusion variance. However, these \acro{AR} models may encounter variance explosion as the prior variances accumulate through the diffusion terms over time, leading to possible overestimation of extremity in later periods \cite{martin2002dynamic}. While one may multiply $\sqrt{1-\frac{\sigma^2_t}{C_i}}$ to $x_{i,t-1}$ to enforce variance balance, information of earlier observations summarized in the prior is discounted. In addition, the implicit Markov assumption impedes utilizing information from future observations in estimating current $x_{i,t}$, as future traits are not represented in the prior. Finally, \acro{AR} trajectories tend to be rough and not well-suited to applications preferring smoother trends.

\section{Proposed method}\label{sec:gdgpirt}

In this section, we present a novel Bayesian approach for dynamic item response theory with ordinal responses. We leverage recent advances in Bayesian non-parametric \acro{IRT} based on Gaussian process (\acro{GP}) for inferring \acro{IRF}s with flexible shapes, and propose a Gaussian process time series model for joint estimation of dynamic latent traits over time with balanced priors. We refer our model as generalized dynamic Gaussian process item response theory (\acro{GD-GPIRT}).

\subsection{Generalized dynamic Gaussian process item response theory}

Our proposed \acro{GD-GPIRT} can be viewed as an extension of the standard 2PL model with an additional Gaussian Process component by assuming that the \acro{IRF}s belongs to a family of Gaussian processes. The additional \acro{GP} part allows non-monotonicity, non-saturation or asymmetry that typically require explicit modeling in other parametric \acro{IRT} \cite{duck2020gpirt}. Mathematically, \acro{GP} is widely used for modeling distributions over functions, such that any realization of functional values has a joint Gaussian distribution. Specifically, \acro{GD-GPIRT} places a hierarchical \acro{GP} prior on each latent $f_{it}$:
\begin{gather}
\label{prior:f}
\mu_{jt}(x) = \beta_{jt1}x + \beta_{jt0},\quad \beta_{jt1}\sim \mathcal{N}(0,\sigma_{\beta_1}^2), 
    \quad \beta_{jt0}\sim \mathcal{N}(0,\sigma_{\beta_0}^2)\\
        K_x(x,x')=\exp\bigl(-\tfrac{1}{2} (x-x')^2 / \ell_x^2 \bigr), \quad p(f_{jt}) \sim \mathcal{GP}(\mu_{jt},K_x)
    \label{kernel:x}
\end{gather}
where each latent function contains an item-specific linear trend $\mu_{jt}(\cdot)$ and a non-linear deviation with a prior $\mathcal{GP}(0,K_x)$. We focus on the unidimensional latent space $\mathcal{X}=\mathbb{R}$ by relying on kernels for implicit transformation to higher dimensions. We use RBF kernel here but our model is not tied to any particular specification of mean and covariance functions and could be carefully chosen either by prior knowledge and/or via Bayesian model selection in practice. The slope and intercept parameters $\boldsymbol{\beta}$s in $\mu_{jt}(\cdot)$ also have zero-mean normal hyper-priors with variance $\sigma_{\beta_1}^2$ and $\sigma_{\beta_0}^2$. The $\ell_x$ parameter is the length scale of the kernel controlling the bandwidth of correlations.

Instead of estimating latent variables at different time periods independently, \acro{GD-GPIRT} models each latent trajectory jointly through time. 
Specifically, we place independent \acro{GP} priors on individual trait vector $\mathbf{x}_{i}=[x_{i1},\dots,x_{iT}]^T$.  Here we use zero mean functions and Mat\'ern kernel of degree $5/2$ to model moderately smooth (twice differentiable) trajectories with length scale $\ell_t$. Hence,  we encode the reasonable expectation that the latent trends are centered and stationary \emph{a priori} Compared to the \acro{AR} model discussed in Sec. \ref{subsec:dynamic_irt}, \acro{GD-GPIRT} ensures measurement comparability as each entry in the dynamic latent trends will have the same marginal prior distribution: 
\begin{gather}
    \label{prior:x}
    p\big(\{\mathbf{x}_{i}\}\big)=\prod_i p(\mathbf{x}_{i}),\quad 
    p(\mathbf{x}_{i})\sim \mathcal{GP}(\mathbf{0},\mathbf{K}_t)\\
    K_t(t,t')=(1+\frac{\sqrt{5} d}{\ell_t}+\frac{5 d^2}{3\ell_t^2})\exp\bigl(-\frac{\sqrt{5} d}{\ell_t}\bigr), d = \lvert t-t' \rvert
\end{gather}

The last set of parameters is the threshold parameters. Following the re-parameterization trick \cite{JMLR:v6:chu05a}, we define a set of positive padding variables $\Delta_l>0$ and $l=2,\dots,C-1$ such that $b_c=b_1+\sum_{l=2}^c\Delta_l$. We place independent standard normal prior on the log scale of the padding variables: $\log(\Delta_l)\sim \mathcal{N}(0,\sigma_{\Delta}^2)$. Note this is equivalent to normal priors on log scale of $b_1, b_2-b_1,\dots,b_{C-1}-b_{C-2}$. Our model parameters include all latent function values $\mathbf{f}_{jt}=[f_{1jt},\dots,f_{njt}]^T$, ability scores $\mathbf{x}_{i}$, slope and intercept parameters $\{\boldsymbol{\beta}_{jt}\}$ and threshold parameters $\{b_{jtc}\}$ with the following joint posterior:
\begin{multline}
\label{eq:jointlik}
    p\big(\{\mathbf{x}_i\},\{\mathbf{f}_{jt}\},\{b_{jtc}\},\{\boldsymbol{\beta}_{jt}\} \mid \{y_{ijt}\} \big)\propto \underbrace{\prod_i p(\mathbf{x}_i)}_{\text{latent trait prior}}
     \underbrace{\prod_{t}\prod_j p(\mathbf{f}_{jt})}_{\text{IRF prior}}
    \underbrace{\prod_{t}\prod_j p(\beta_{jt1})}_{\text{slope prior}} \\
     \underbrace{\prod_{t}\prod_j p(\beta_{jt0})}_{\text{intercept prior}}
    \underbrace{\prod_{t}\prod_j\prod_{c\vphantom{j}} p(b_{jtc})}_{\text{threshold prior}}
    \underbrace{\prod_{t}\prod_{i}\prod_j p\big( \{y_{ijt}\} \mid \{\mathbf{x}_i\},\{\mathbf{f}_{jt}\},\{b_{jtc}\},\{\boldsymbol{\beta}_{jt}\} \big)}_{\text{likelihood}}
\end{multline}

\subsection{Model inference}
\label{sec:inference}
We propose a Markov chain Monte Carlo (\acro{MCMC}) sampler for posteriors of latent variables and \acro{IRF}s, as they are highly multivariate. We proceed in a Gibbs sampling fashion. Let superscripts $k$ denote the $k$th iteration in the \acro{MCMC} sampler. The sampler is initialized by drawing $\{\mathbf{x}_{i}^{(0)}\}$s, $\{\boldsymbol{\beta}^{(0)}_{jt}\}$s and $\{b^{(0)}_c\}$s from their respective priors, and drawing the latent function values $\{\mathbf{f}^{(0)}_{jt}\}$s from the induced multivariate Gaussian at all $n$ initial locations $\mathbf{x}_t^{(0)}$ at time $t$:
\begin{equation}
\label{initialize:f}
p\big(\mathbf{f}^{(0)}_{jt}\mid \mathbf{x}^{(0)}_t,\boldsymbol{\beta}^{(0)}_{jt}\big)\sim\mathcal{GP}\big(\mu_{jt}(\mathbf{x}^{(0)}_t; \boldsymbol{\beta}^{(0)}_{jt}), \textbf{K}_x(\mathbf{x}^{(0)}_{t},\mathbf{x}^{(0)}_{t})\big)
\end{equation}
After initialization, we alternatively sample each variable in the targeted joint distribution in Eq. (\ref{eq:jointlik}) from its conditioning distribution on all the other variables. First, the sampler draws new latent function values by conditioning on all $\{y_{ijt}\}$s:
\begin{equation}
\label{sample:f}
p\big(\mathbf{f}^{(k+1)}_{jt}\mid \mathbf{x}^{(k)}_{t}, \{y_{ijt}\},\boldsymbol{\beta}^{(k)}_{jt},\{b^{(k)}_c\}\big)  \propto  p\big(\mathbf{f}^{(k+1)}_{jt}\mid \mathbf{x}^{(k)}_{t}, \boldsymbol{\beta}^{(k)}_{jt} \big)
     p\big(\{y_{ijt}\}\mid \mathbf{f}^{(k+1)}_{jt},\{b^{(k)}_c\}\big)
\end{equation}
As \acro{GP} regression allows no analytical form for non-Gaussian likelihood, we exploit eclipse slice sampling (\acro{ESS}) \cite{murray2010elliptical} for the conditional distributions. Eclipse slice sampling is a generic sampler for posterior of arbitrary target variable $\mathbf{z}$ with a Gaussian prior $\mathbf{z}\sim \mathcal{N}(\boldsymbol{\mu},\boldsymbol{\Sigma})$ and a likelihood function $\mathcal{L}(\mathbf{z})$, and more efficient than the traditional Metropolis–Hastings stepping. \acro{ESS} samples the next iteration by adaptively performing slicing sampling on the eclipse defined by current state $\mathbf{z}$ and a random draw $\boldsymbol{\nu}$ from prior $\mathcal{N}(\boldsymbol{\mu},\boldsymbol{\Sigma})$ (see Supplement for details). The prior mean $\boldsymbol{\mu}_f$, covariance $\boldsymbol{\Sigma}_f$ and likelihood $\mathcal{L}(\mathbf{f_{jt}})$ for sampling $\mathbf{f}^{(k+1)}_{jt}$ are defined as:
\begin{gather}
\boldsymbol{\mu}_f = \mu_{jt}\big(\mathbf{x}^{(k)}_{t};\boldsymbol{\beta}^{(k)}_{jt}\big), \quad
\boldsymbol{\Sigma}_f=K_x\big(\mathbf{x}^{(k)}_{t},\mathbf{x}^{(k)}_{t}\big)\\
\mathcal{L}(\mathbf{f}_{jt}) = \prod_{i}p\big( \{y_{ijt}\} \mid \mathbf{x}_t,\{{f}_{ijt}\},\{b_c\},\boldsymbol{\beta}_{jt} \big)
\end{gather}
After obtaining $\mathbf{f}^{(k+1)}_{jt}$s for all items at all time periods, we then sample the latent trait $\mathbf{x}^{(k+1)}_{i}$s. However, direct sampling from the conditional distribution of $\mathbf{x}^{(k+1)}_{i}$ is not obvious, because the likelihood of locations is not defined by $\mathbf{f}_{jt}$s at points other than $\mathbf{x}^{(k+1)}_{i}$s. Hence, we introduce a set of auxiliary variables $\mathbf{f}^*_{jt}$, which are the latent functions values defined on an evenly-spaced dense grid $\mathbf{x}^{*}$ from $-5$ to $5$ in one-hundred increment. We used $500$ inducing points here as shown sufficient for \acro{GP}-based latent variable modeling \citep{uhrenholt2021probabilistic}. Samples of $\mathbf{f}^*_{jt}$ can be obtained by applying \acro{GP} posterior update rule, conditioning on current location $\mathbf{x}^{(k)}_{t}$s and function value $\mathbf{f}^{(k+1)}_{jt}$s:
\begin{gather}
\label{ess:f}
    p\big( \mathbf{f^*}^{(k+1)}_{jt} \big)\sim\mathcal{GP}(\boldsymbol{\mu}^*,K^*),\quad \boldsymbol{\mu}^*=K_x\big(\mathbf{x}^{*},\mathbf{x}^{(k)}_{t}\big)\mathbf{V}^{-1}\mathbf{f}^{(k+1)}_{jt}\\
    K^*=K_x\big(\mathbf{x}^{*},\mathbf{x}^{*}\big)-K_x\big(\mathbf{x}^{*},\mathbf{x}^{(k)}_{t}\big)\mathbf{V}^{-1}K_x\big(\mathbf{x}^{(k)}_{t},\mathbf{x}^{*}\big),\quad \mathbf{V} = K_x\big(\mathbf{x}^{(k)}_{t},\mathbf{x}^{(k)}_{t}\big)
\end{gather}
Note that the covariance matrix $\mathbf{V}$ is the same for all question $j$s so its inversion only needs to be computed once. With these auxiliary variables $\mathbf{f}^*_{jt}$, we obtain a dense approximation of likelihood values for all latent locations besides $\mathbf{x}^{(k)}_{t}$. We construct mean $\boldsymbol{\mu}_x$, covariance $\boldsymbol{\Sigma}_x$ and likelihood $\mathcal{L}(\mathbf{x}_{i})$ for sampling $\mathbf{x}^{(k+1)}_{i}$ with time vector $\mathbf{t}_i=[1,\dots,T]^T$ as:
\begin{equation}
\label{ess:x}
\boldsymbol{\mu}_x=\mathbf{0}, \quad \boldsymbol{\Sigma}_x=K_t\big(\mathbf{t}_i,\mathbf{t}_i\big),\quad 
\mathcal{L}(\mathbf{x}_{i}) = \prod_{j}\prod_{t}p\big( \{y_{ijt}\} \mid \mathbf{x}_i,\{\mathbf{f}^*_{jt}\},\{b_c\},\boldsymbol{\beta}_{jt} \big)
\end{equation}
The latent trait location samples are then rounded to the nearest rug in the dense grid $\mathbf{x}^{*}$. We further update the latent function values $\mathbf{f}^{(k+1)}_{jt}$ to those $\mathbf{f^*}^{(k+1)}_{jt}$ defined on new $\mathbf{x}^{(k+1)}_{t}$. Finally, we sample new slope and intercept parameters $\{\boldsymbol{\beta}^{(k+1)}_{jt}\}$s and threshold parameters $\{b^{(k+1)}_c\}$s using \acro{ESS}, based on the new latent locations $\mathbf{x}^{(k+1)}_{t}$ and updated function values $\mathbf{f}^{(k+1)}_{jt}$.

Our inference procedure can also be further adjusted when the latent item function $\mathbf{f}_{jt}$s come from the same set of items, meaning $\mathbf{f}_{j1}=\dots=\mathbf{f}_{jt}=\mathbf{f}_{j}$ for all $j$s. Now inference of $\mathbf{f}_{j}$ need to condition on all $nT$ latent traits $\{{x}_{it}\}$s and corresponding observations $\{y_{ijt}\}$s, making sampling of auxiliary $\mathbf{f^*}_{j}$s computationally demanding. Hence, we exploit a sparse GP trick that selects $100$ inducing locations on the dense grid whose inducing values are determined by its $k$-nearest neighbors. In our exploration, we found notable speed-up when $nT\approx 6,000$ but no performance lost.

\section{Experiments}\label{sec:experiments}

As latent variables are not directly observable, we use simulation to analyze measurement and prediction quality of \acro{GD-GPIRT}. We also apply \acro{GD-GPIRT} with two real-world case studies and examine  the forecasting ability of \acro{GD-GPIRT}.

\subsection{Simulation studies}

\paragraph{Data generating process:} The simulation 
consists of $100$ synthetic respondents and $10$ items over $10$ time periods. We consider two scenarios with binary ($C=2$) or ordinal ($C=5$) responses with different sets of items across time. The latent vectors are drawn i.i.d from the zero-mean \acro{GP} defined in Eq. (\ref{prior:x}) with $\ell_t=5$, and the \acro{IRF}s from the \acro{GP} in Eq. (\ref{kernel:x}) with $\ell_x=1$ and $\sigma^2_{\beta_1}=\sigma^2_{\beta_1}=1$. We also draw $\{b_c\}$s from $\text{Unif}[-2,2]$, and sort $\{b_c\}$s to ensure all ordinal responses have non-zero probabilities. Finally, we generate responses $y_{ij}$ from the probabilistic model defined in Eq. (\ref{likelihood}).

\begin{table}[t!]
\begin{center}
    \caption{{\small Comparison of measurement quality and predictive fit between \acro{GD-GPIRT} and baselines under two synthetic settings. Bold numbers indicate statistical significance compared to the other methods using standard paired t-tests and italicized numbers indicate the method is not statistically worse than the best method. Overall, \acro{GD-GPIRT} performs consistently better over baselines in terms of the measurement quality of estimated traits and \acro{IRF}s while having better prediction of the actual responses.}}
\label{table:simulation}
\resizebox{\linewidth}{!}{
\begin{tabular}{lrrrrrr}
\toprule
& \multicolumn{3}{c}{\bftab Binary Response ($C=2$)}  & \multicolumn{3}{c}{ \bftab Ordinal Response ($C=5$)}  
\\
\cmidrule{2-7}
\acro{MODEL} & $\acro{cor}(\mathbf{x},\mathbf{\hat{x}}) \boldsymbol{\uparrow}$ & $\acro{RMSE}( \text{ICC}) \boldsymbol{\downarrow}$  & $\acro{ACC}(y,\hat{y})\boldsymbol{\uparrow}$ & $\acro{cor}(\mathbf{x},\mathbf{\hat{x}})\boldsymbol{\uparrow}$ & $\acro{RMSE}( \text{ICC})\boldsymbol{\downarrow}$  & $\acro{ACC}(y,\hat{y})\boldsymbol{\uparrow}$  \\
\midrule
\acro{ggum} & 0.433$\pm$0.011   & 0.435$\pm$0.001  & 0.478$\pm$0.010 &  0.531$\pm$0.013  &  0.654$\pm$0.027 &  0.205$\pm$0.005\\ 
\acro{grm} &  0.468$\pm$0.012  & 0.374$\pm$0.005  & 0.583$\pm$0.005 &  0.545$\pm$0.014  &  0.407$\pm$0.015 &  0.235$\pm$0.003\\ 
\acro{gpcm} &  0.486$\pm$0.014  & 0.379$\pm$0.004  & 0.587$\pm$0.005 &  0.550$\pm$0.012  & 0.350$\pm$0.019  & 0.303$\pm$0.003\\ 
\acro{srm} & 0.492$\pm$0.012   & 0.399$\pm$0.005 & 0.522$\pm$0.007 &   0.543$\pm$0.013 &  0.580$\pm$0.018 & 0.243$\pm$0.003\\ 
\acro{dsem} & 0.486$\pm$0.017   & 0.431$\pm$0.005 & 0.509$\pm$0.017 &   0.594$\pm$0.012 &  0.587$\pm$0.048 & 0.214$\pm$0.007\\ 
\midrule
\acro{GPDM} & \text{0.485$\pm$0.018}  & \text{0.306}$\pm$0.001  & \text{0.647}$\pm$0.007 & \text{0.434$\pm$0.004} & 
\text{0.339$\pm$0.008}  & \text{0.178}$\pm$0.011 \\
\acro{GPIRT} & \text{0.855$\pm$0.005}  & \text{0.103}$\pm$0.003  & \text{0.708}$\pm$0.010 & \text{0.927$\pm$0.004} & 
\textit{0.272$\pm$0.008}  & \text{0.285}$\pm$0.011 \\
\acro{L-GPIRT} &  \text{0.878$\pm$0.005} &  \text{0.102$\pm$0.003} & \text{0.674$\pm$0.009} & \text{0.867$\pm$0.008} & \text{0.444$\pm$0.035}   & \text{0.248$\pm$0.008} \\
\acro{DO-IRT} &  0.896$\pm$0.003   & 0.148$\pm$0.002 & 0.751$\pm$0.007 &  \text{0.920}$\pm$0.002 &  \text{0.449}$\pm$0.006 & 0.418$\pm$0.008 \\ 
\acro{GD-GPIRT} & \textbf{0.961$\pm$0.005} &  \textbf{0.089$\pm$0.003}  &  \textbf{0.794$\pm$0.006} & \textbf{0.981$\pm$0.002} & \textbf{0.262$\pm$0.012}  &  \textbf{0.455$\pm$0.011} \\
\bottomrule
\end{tabular}}
\end{center}
\end{table}                                                        

\paragraph{Baselines and metrics:} We compare \acro{GD-GPIRT} to (1) several ordinal \acro{IRT} models, including the generalized graded unfolding model (\acro{GGUM}), the graded response model (\acro{GRM}) \citep{samejima1969estimation}, the generalized partial credit model (\acro{GPCM}) \citep{muraki1992generalized} and the sequential response model (\acro{SRM}) \citep{tutz1990sequential}, (2) dynamic structural equation model (\acro{DSEM}) \citep{asparouhov2018dynamic, mcneish2023dynamic}, and (3) the dynamic ordinal \acro{IRT} (\acro{DO-IRT}) model with auto-regressive Wiener latent trends \cite{schnakenberg_fariss_2014}. We also consider baselines from \acro{GPLVM} literature, including Gaussian process dynamic model (\acro{GPDM}) \citep{damianou2011variational}, the static Gaussian process \acro{IRT} model (\acro{GPIRT}) \citep{duck2020gpirt} that estimates each time period separately, and an ablation model  without non-linear deviation in the latent process (\acro{L-GPIRT}). We focus on three metrics for evaluating measurement quality of estimated latent traits and \acro{IRF}s as well as predictive fit of responses: 1) the averaged normalized correlation of the estimated traits; 2) the \acro{RMSE} of item characteristics curves (\acro{ICC}), or the expected response $\text{ICC}(x;f,\mathbf{b})=\mathbb{E}[y\mid f,\mathbf{b},x]$; and 3) the predictive accuracy $\acro{ACC}(y,\hat{y})$ of actually responses.

\paragraph{Results:} We split data into 80\%/20\% for training and testing. We repeat each simulation setting using $25$ different seeds with 300 Intel Xeon 2680 CPUs. For each run, we simulate three \acro{MCMC} chains with $500$ burnout steps and $500$ sampling iterations, thinned every four samples. With training data size of $10k$, our software managed to finish all $1k$ iterations in around 15 minutes. Our sampler also demonstrates favorable convergence, with all R-hat diagnostics below $1.1$ and effective sample sizes exceeding 100 in all runs (see Appendix for details). Table \ref{table:simulation} shows comparison of measurement quality and predictive fit between \acro{GD-GPIRT} and baselines under two synthetic settings. Bold numbers indicate statistical significance compared to the other methods using standard paired t-tests and italicized numbers indicate the method is not statistically worse than the best method. Overall, \acro{GD-GPIRT} performs consistently better over baselines in terms of the measurement quality of estimated traits and \acro{IRF}s while having better prediction of the actual responses.

\subsection{Public opinions on economy}

\paragraph{Data:} The American Panel Survey (\acro{TAPS}) was a long-running research project to study public opinions from all 50 U.S. states and included an extensive array of survey items asked across multiple waves. Specifically, respondents were asked questions monthly from Jan. 2014 to Jan. 2018 about their opinions on the economic conditions and income allocation (either to spend or to save) of their households and the country as a whole. Since the same set of questions were repeatedly asked, we apply \acro{GD-GPIRT} with sparse \acro{GP} speedup to estimate people's financial status and how they react to those questions. We set $\ell_t=12$ to capture yearly shifts in attitudes.

\begin{figure}
  \centering
  \begin{subfigure}[b]{.6\linewidth}
      {\includegraphics[width=\linewidth]{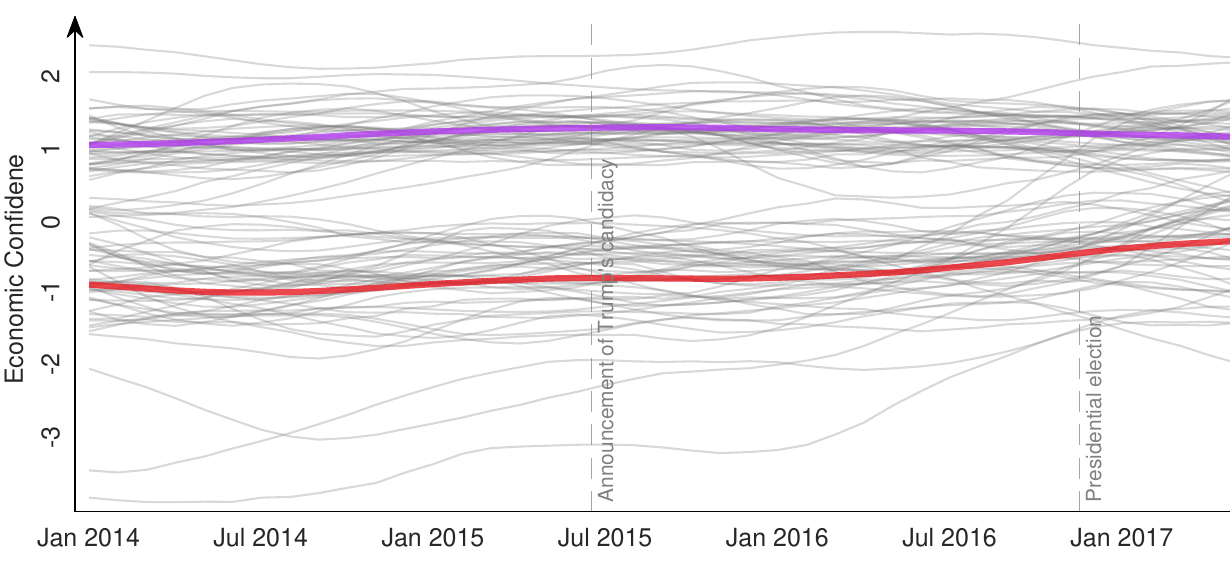}}
      \caption{\small{Illustration of public opinion trends in economic confidence. The grey lines represent our estimated individual confidence levels estimated, while the bold blue and red lines depict the average confidence levels for Democrats and Republicans respectively. Vertical dashed lines mark key events such as the announcement of Trump's candidacy and the 2016 Presidential election. Notably, Dem. levels remained relatively stable, while Rep. exhibited a slight increase in confidence leading up, following the election of Trump.}}
    \label{fig:gpirt_TAPS_scores}
  \end{subfigure}%
  \hfill 
  \begin{subfigure}[b]{.4\linewidth}
  \begin{subfigure}[t]{\textwidth}
        \includegraphics[width=\textwidth]{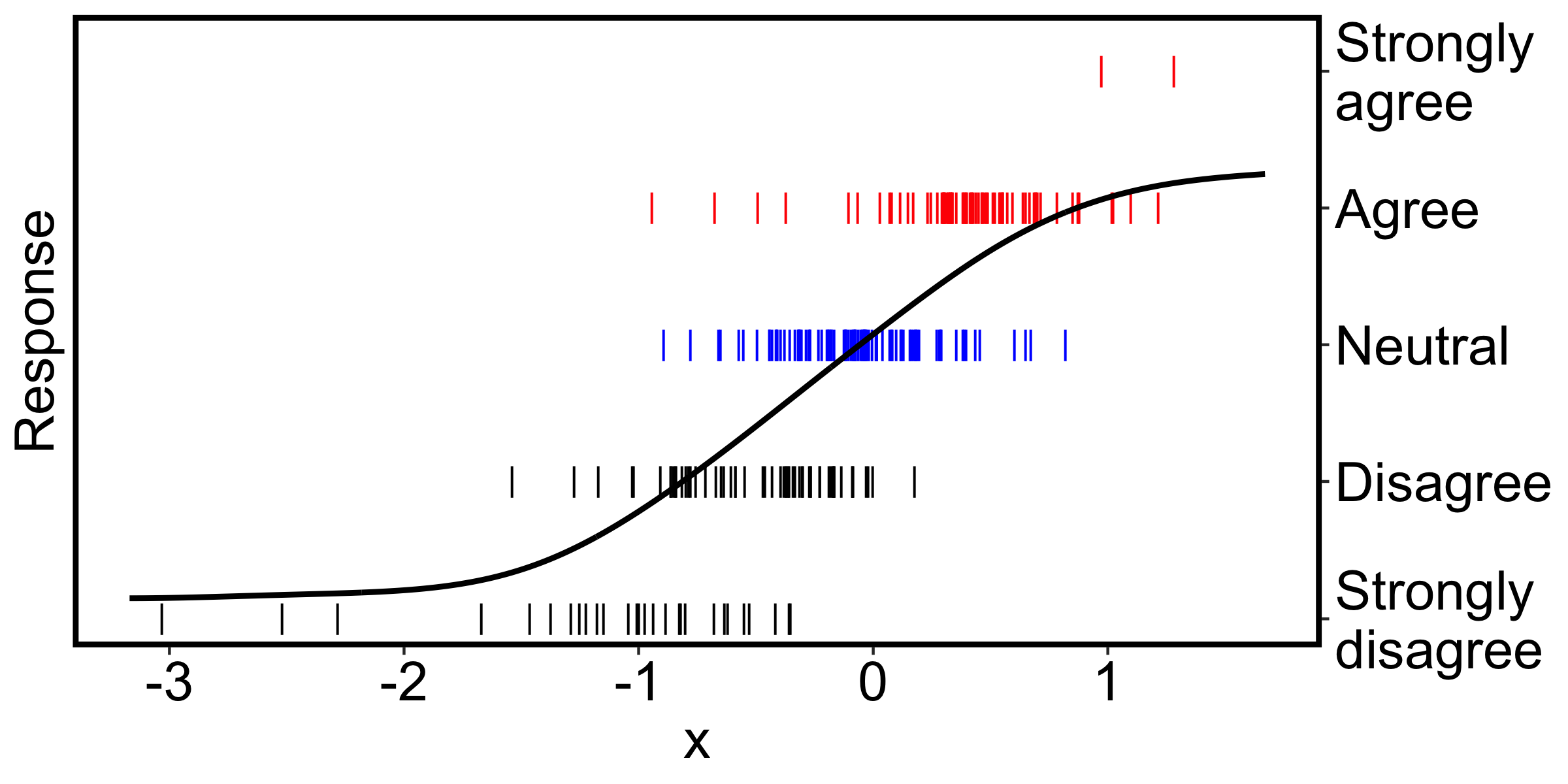}
        \caption{\scriptsize{Economic conditions in U.S. are getting better.}}
      \end{subfigure}

      \begin{subfigure}[b]{\textwidth}
        \includegraphics[width=\textwidth]{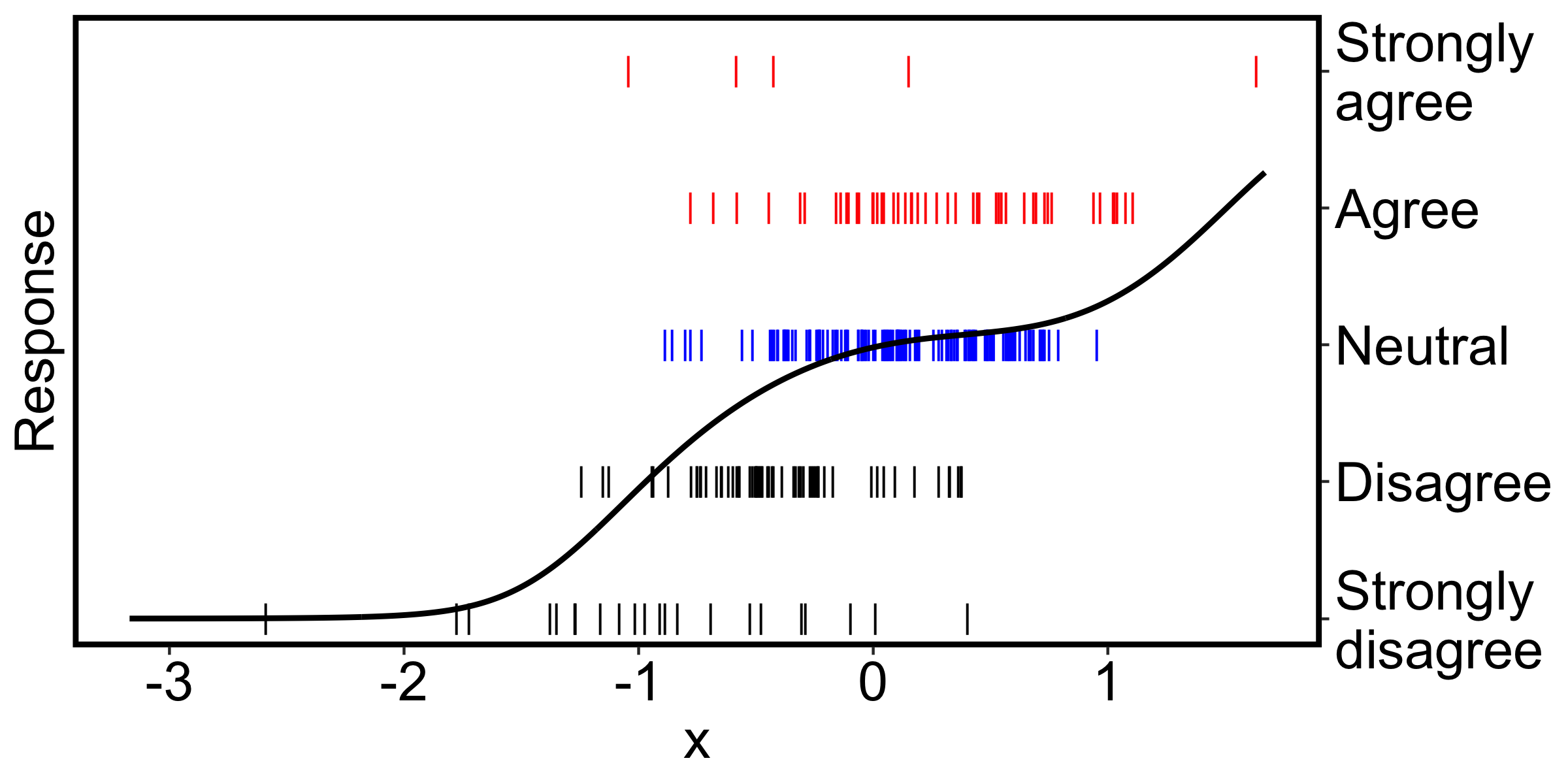}
        \caption{\scriptsize{Compared to past few months, I expect to spend.}}
      \end{subfigure}
  \end{subfigure}%
  \caption
    {{\small Estimated confidence and selective \acro{IRF}s in \acro{TAPS}.}}
    \label{fig:TAPS}
\end{figure}



\paragraph{Estimated economic confidence:} Figure \ref{fig:gpirt_TAPS_scores} illustrates public opinion trends in economic confidence. Grey lines represent individual confidence levels estimated using \acro{GD-GPIRT}, while bold blue and red lines depict averaged confidence levels among Democrats and Republicans. Vertical dashed lines mark key events such as the announcement of Trump's candidacy and the 2016 Presidential election. Notably, Democrats' confidence levels remained relatively stable, while Republicans exhibited a slight increase in confidence leading up to and following the election of President Trump. This is consistent with existing theories on partisan sources of confidence. Selective estimated \acro{IRF}s is shown in Figure \ref{fig:TAPS}. In general, as economic confidence diminishes, respondents are more inclined to disagree that economic conditions are improving, and they have reduced expectations for savings.

\begin{table}[H]
\centering
\caption{{\small Predictive \acro{ACC} and log lik of \acro{GD-GPIRT} and baselines in predicting future responses at various forecasting horizons. Italicized numbers indicate methods that are not statistically worse than the best method. \acro{GD-GPIRT} significantly outperforms all baselines for all forecasting horizons expect for \acro{GPDM} at far horizons. }}
\label{table:tapsforecast}
\resizebox{\linewidth}{!}{
\begin{tabular}{lrrrrrr}
\\
\toprule
 & \multicolumn{3}{c}{\bftab \acro{Acc} $\boldsymbol{\uparrow}$} & \multicolumn{3}{c}{\bftab \acro{LL} $\boldsymbol{\uparrow}$} \\
\cmidrule{2-7} 
\bftab \acro{Horizon} & $1$ month & $6$ months & $12$ months & $1$ month & $6$ month & $12$ month \\
\midrule
\acro{ggum}  & \text{0.182$\pm$0.003} & \text{0.150$\pm$0.003} & \text{0.158$\pm$0.003} & \text{ $-$12.74$\pm$0.085} & \text{ $-$17.58$\pm$0.155} & \text{ $-$13.07$\pm$0.100}\\
\acro{grm}  & \text{0.289$\pm$0.005} & \text{0.300$\pm$0.006} & \text{0.244$\pm$0.005} & \text{ $-$4.435$\pm$0.037} & \text{ $-$4.864$\pm$0.044} & \text{ $-$4.737$\pm$0.035}\\
\acro{gpcm}  & \text{0.408$\pm$0.007} & \text{0.382$\pm$0.008} & \text{0.397$\pm$0.007} & \text{ $-$1.364$\pm$0.019} & \text{ $-$1.584$\pm$0.027} & \text{ $-$1.404$\pm$0.016}\\
\acro{srm}  & \text{0.298$\pm$0.005} & \text{0.388$\pm$0.006} & \text{0.279$\pm$0.006} & \text{ $-$2.137$\pm$0.026} & \text{ $-$1.876$\pm$0.029} & \text{ $-$2.249$\pm$0.028}\\
\acro{dsem}  & \text{0.215$\pm$0.010} & \text{0.309$\pm$0.009} & \text{0.288$\pm$0.007} & \text{ $-$1.979$\pm$0.039} & \text{ $-$1.658$\pm$0.041} & \text{ $-$1.692$\pm$0.023}\\
\midrule
\acro{GPDM}  & \text{0.552$\pm$0.007} & \text{ 0.600$\pm$0.009} & \textit{0.558$\pm$0.007} & \text{ $-$1.058$\pm$0.014} & \textit{ $-$0.924$\pm$0.009} & \textit{ $-$1.099$\pm$0.012}\\
\acro{GPIRT}  & \text{0.526$\pm$0.006} & \text{0.510$\pm$0.007} & \text{0.546$\pm$0.008} & \text{ $-$1.282$\pm$0.017} & \text{ $-$1.124$\pm$0.015} & \text{ $-$1.198$\pm$0.021}\\
\acro{L-GPIRT}  & \text{0.488$\pm$0.025} & \text{0.493$\pm$0.035} & \text{0.470$\pm$0.023} & \text{ $-$1.222$\pm$0.053} & \text{ $-$1.200$\pm$0.066} & \text{ $-$1.287$\pm$0.041}\\
\acro{DO-IRT}  & \text{0.573$\pm$0.006} & \text{0.604$\pm$0.007} & \text{0.538$\pm$0.007} & { $-$1.144$\pm$0.025} & \textit{ $-$0.957$\pm$0.021} & { $-$1.276$\pm$0.024} \\
\bftab \acro{GD-GPIRT} & {\bftab 0.597$\pm$0.005} & {\bftab 0.665$\pm$0.008} & {\bftab 0.572$\pm$0.005} & {\bftab $-$1.003$\pm$0.008} & \textbf{ $-$0.916$\pm$0.008} & \textbf{ $-$1.074$\pm$0.013}  \\
\bottomrule
\end{tabular}
}
\end{table}

\paragraph{Forecasting future responses:} In order to assess the  out-of-sample prediction of \acro{GD-GPIRT} with respect to future responses, we conducted a forecasting analysis. Specifically, we first estimate confidence levels of each individual based on data spanning from 2014 to 2017, and then extrapolate their confidence levels in 2018. We then hold out 20\% of distinct individuals for every question, leverage the remaining 80\% of observations to estimate \acro{IRF}s and predict their future responses from the extrapolated confidence levels across multiple forecasting horizons ranging from 1 to 12 months. Table \ref{table:tapsforecast} shows the predictive accuracy and log likelihood of \acro{GD-GPIRT} and baselines in forecasting future responses at various horizons. Our findings show that \acro{GD-GPIRT} significantly outperforms all baselines for all forecasting horizons expect for \acro{GPDM} at far horizons, suggesting effectiveness of \acro{GD-GPIRT} in modeling the trajectories of confidence levels and its measurement quality.

\subsection{Ideology of U.S. Senate on abortion debate}

\paragraph{Data:} Ideology \cite{converse1964}, plays a central role in understanding congressional dynamics such as political polarization and partisan sorting \cite{poole2001dwnominate, fiorina2008polarization}. Scaling congressional votes to ideology faces challenges in accommodating temporal changes while ensuring comparability, as politicians have demonstrated substantial shifts in their views over time \cite{biden2019, biden2022}. However, previous studies often simplify complex ideology trends with linear models or low order polynomials \cite{poole2001dwnominate, bailey2007comparable, bailey2013today}. We run \acro{GD-GPIRT} to estimate the U.S.\ Senate's ideology over multiple congressional sessions, using roll-call voting data obtained from the \textit{voteview} database \cite{lewis2019voteview}. We focus on votes related to abortion debate as identified in \cite{Montgomery:2011}, resulting in 758k total votes of 235 Senators spanning over 20 years. We set $\ell_t=6$, the maximum-a-posterior estimate from a simple \acro{GP} model of \acro{Nokken--Poole} scores \cite{nokken2004congressional} against time.

\begin{figure}[H]
   \hspace*{-2em}\centerline{\includegraphics[width=\linewidth]{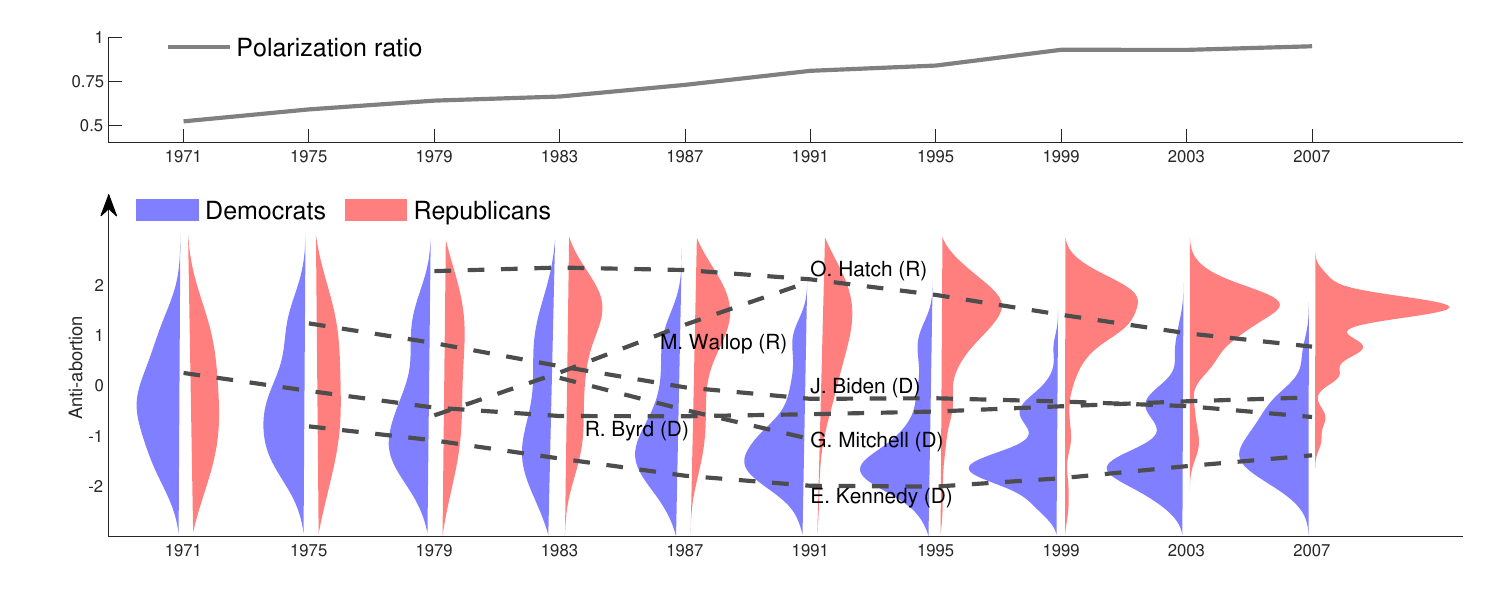}}
     \caption{\small{The upper panel illustrates the alignment between senators' estimated ideology and their party affiliations ('polarization ratio') on abortion debates using \acro{GD-GPIRT}. The increase in the ratio from the 92nd Congress (0.521) to the 108th Congress (0.950) indicates the growing partisan divide on abortion debate. The lower panel displays \acro{GD-GPIRT} scores by party spaced every two sessions, with Dem. and Rep. Senators in blue and red, respectively. Dashed lines represent the evolving ideological trajectories of selected senators.}}
    \label{fig:abortionscore}
\end{figure}

\begin{figure}[H]
        \begin{subfigure}[b]{0.24\linewidth}
             \includegraphics[width=\linewidth]{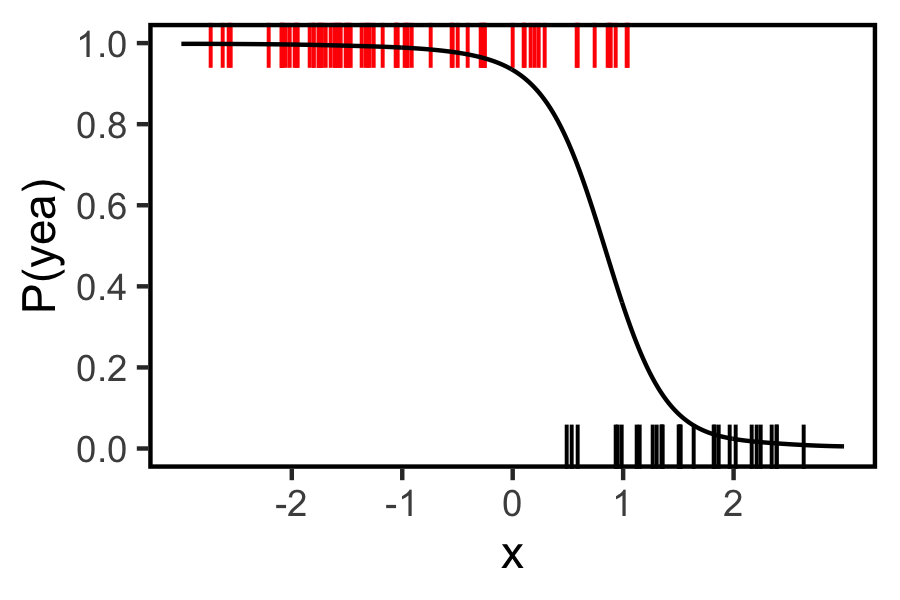}
                \caption{{\small Public Health Service Act to permit free access to medical facilities.}} 
                \label{fig:abortionirf1}
        \end{subfigure}\hfill
        \begin{subfigure}[b]{0.24\linewidth}
             \includegraphics[width=\linewidth]{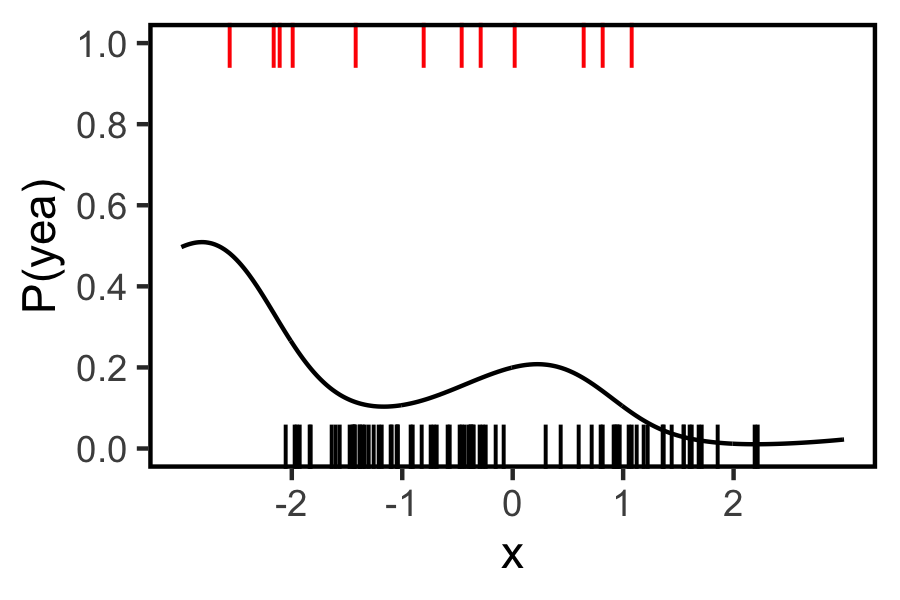}
                \caption{{\small Authorization of extra \$5M funds for family planning project grants.}}
                \label{fig:abortionirf2}
        \end{subfigure}\hfill
        \begin{subfigure}[b]{0.24\linewidth}
              \includegraphics[width=\linewidth]{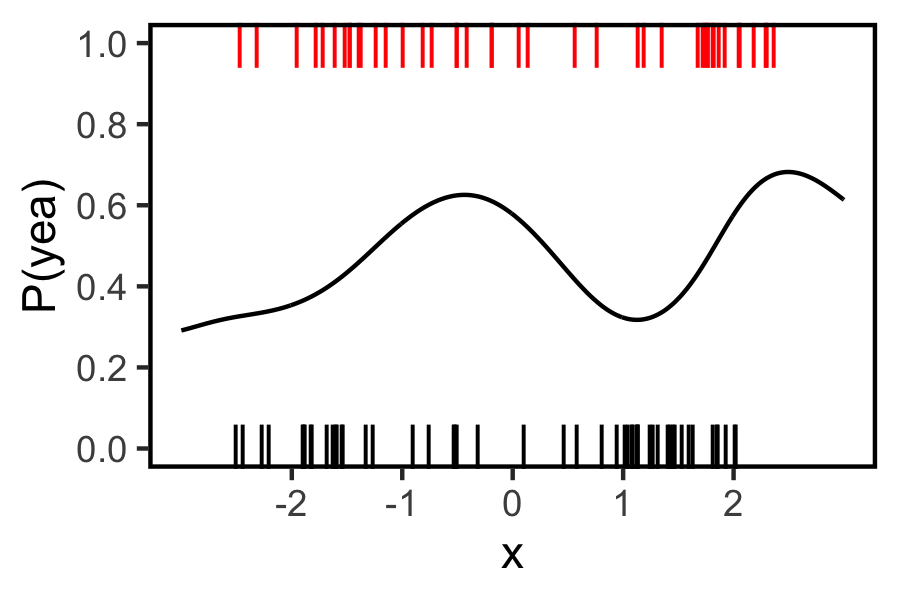}
                \caption{{\small Requirement of monitoring regarding forced abortions in China.}}
                \label{fig:abortionirf3}
        \end{subfigure}\hfill
        \begin{subfigure}[b]{0.24\linewidth}
 \includegraphics[width=\linewidth]{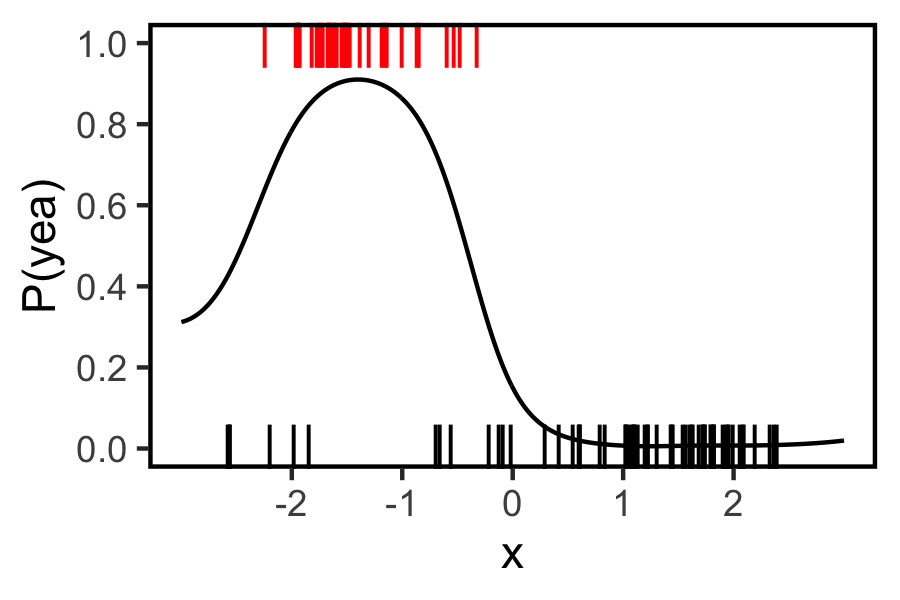}
                \caption{{\small Prohibition of abortion where fetus is determined to be viable.}}
                \label{fig:abortionirf4}
        \end{subfigure}
        \caption{{\small \acro{IRF}s of four selected roll call votes in the U.S.\ Senate on abortion debate between the 92th to 108th Congress that are standard, asymmetric, non-saturate and non-monotonic. Estimated probability of voting ``yea'' is plotted against ideology score $x$. Actual ``yea'' and ``nay'' roll-call votes are displayed as red and black dashes.}}
        \label{fig:abortionirf}
\end{figure}

\paragraph{Increased polarization:} Our results reveal a clear and noticeable pattern of increasing partisan polarization within the United States Congress on abortion debate. In Figure \ref{fig:abortionscore}, the upper panel illustrates the alignment between senators' estimated ideology and their party affiliations, denoted as 'polarization ratio', on abortion debate under \acro{GD-GPIRT}. The increase in the ratio from the 92nd Congress (0.521) to the 108th Congress (0.950) indicates the growing partisan divide on abortion debate. The lower panel displays \acro{GD-GPIRT} scores by party spaced every two sessions, with Dem. and Rep. Senators in blue and red, respectively. Dashed lines represent the evolving ideological trajectories of selected senators. The prominent party divide emerging in the 1990s underscores the partisan sorting related to abortion debate, aligned with prior research \cite{brady1995ideology, adams1997abortion}. 
\paragraph{Non-standard response functions.} Furthermore, \acro{GD-GPIRT} is capable of inferring \acro{IRF}s beyond standard logistic shapes. Figure \ref{fig:abortionirf} shows \acro{IRF}s for four selected roll call votes in the US Senate between the $92$th to $108$th Congress. The estimated probability of voting ``yea'' is plotted against the ideology score $x$. The actual ``yea'' and ``nay'' roll-call votes are displayed as red and black vertical dashes. From left to right, these \acro{IRF}s are either standard, asymmetric, non-monotonic, or non-saturate (do not approach zero or one). 

\paragraph{Vote prediction:} We also assess the predictive performance of \acro{GD-GPIRT} regarding actual responses. Table \ref{table:abortion} shows the comparison of model fits between \acro{GD-GPIRT} and baselines in \acro{TAPS} and \acro{Congress} applications. \acro{GD-GPIRT} has significant higher prediction accuracy, log likelihood and area under ROC curve (AUC) than all \acro{IRT} and \acro{GPLVM} baselines.

\begin{table}[ht]
\caption{{\small Comparison of model fits between \acro{GD-GPIRT} and baselines in \acro{TAPS} and \acro{Congress} applications. \acro{GD-GPIRT} has significant higher prediction accuracy, log lik and AUC than all \acro{IRT} and \acro{GPLVM} baselines.}}
\label{table:abortion}
\begin{center}
{\begin{tabular}{lcccccc}
\toprule
& \multicolumn{2}{c}{\bftab $\mathcal{L}/N$ $\boldsymbol{\uparrow}$} & \multicolumn{2}{c}{\bftab  \acro{ACC} $\boldsymbol{\uparrow}$} & \multicolumn{2}{c}{\bftab \acro{AUC} $\boldsymbol{\uparrow}$} \\
\cmidrule{2-7}
\bftab \acro{data} & \acro{TAPS} & abortion & \acro{TAPS} & \acro{Congress} & \acro{TAPS} & \acro{Congress} \\
 \midrule
 \acro{ggmu} & $-$1.240 & $-$1.355 & 0.353  & 0.475 & 0.365 & 0.493 \\
 \acro{grm} & $-$4.631 &  $-$1.314 & 0.271 & 0.330 &  0.275 & 0.584 \\
 \acro{gpcm} & $-$1.016 &  $-$0.549 & 0.532 & 0.667 & 0.540 &  0.664 \\
 \acro{srm} & $-$2.663 &  $-$1.168 & 0.234 & 0.363 & 0.241 & 0.640  \\
 \acro{DSEM} & $-$1.530 &  $-$1.150 & 0.402 & 0.560 & 0.406 & 0.543 \\
 \midrule
 \acro{GPDM} & $-$0.999 &  $-$0.563 & 0.558  & 0.759 & 0.616 & 0.757\\
\acro{GPIRT} & $-$1.013 &  $-$0.602 & 0.571  & 0.668 & 0.665 & 0.667\\
\acro{L-GPIRT} & $-$1.024 &  $-$0.557 & 0.586  & 0.733 & 0.653 & 0.730 \\
\acro{DO-IRT} & $-$0.979 &  {$-$0.402} & 0.593  & {0.825} & 0.630 & {0.818}  \\
{\bftab \acro{GD-GPIRT}} & \bftab $-$0.949  & \text{\bftab $-$0.160} & \bftab {0.600}  & \text{\bftab 0.930} & \bftab {0.687} & \text{\bftab 0.930}  \\
\bottomrule
\end{tabular}}
\end{center}
\end{table}

\section{Conclusion}\label{sec: conclusion}

We propose a generalized dynamic Gaussian process item response theory (\acro{GD-GPIRT}) for latent measurement, for estimating dynamic latent traits while making minimal assumptions about shapes of the response functions. We validate the measurement quality and prediction of \acro{GD-GPIRT} using simulation studies. We also apply \acro{GD-GPIRT} to address two real-world problems, and show \acro{GD-GPIRT} has better forecasting of future responses than existing methods from both \acro{IRT} or \acro{GPLVM} literature.

We see potentials of \acro{GD-GPIRT} in the advancement of \acro{IRT} in several ways. \acro{GD-GPIRT} can be extended to model multi-dimensional latent traits, which is particularly relevant in fields such as political science where traits like the Big Five personality traits \cite{gerber2011big} and the 2-d \acro{NOMINATE} scores \cite{poole2001dwnominate} are essential. In addition, by clustering models such as mixtures of \acro{GP}s with Dirichlet process prior, the conditional independence assumption among individual traits may be further relaxed when certain participating units naturally form subgroups. Finally, contextual information such as behaviors or demographics could also be incorporated in the kernel to allow more precision latent trait estimation.


\begin{ack}
YC and RG were supported by the National Science Foundation (\acro{NSF}) under award number \acro{IIS}–1845434.
\end{ack}

\bibliography{reference}
\bibliographystyle{unsrtnat}


\newpage
\appendix

\section{Summary of Elliptical Slice Sampling}

Alg. (\ref{alg:ess}) summarizes the elliptical slice sampling.
\begin{algorithm}
\caption{Elliptical slice sampling from drawing new posterior $\mathbf{z'}\sim \mathcal{N}(\boldsymbol{\mu},\boldsymbol{\Sigma})\mathcal{L}(\mathbf{z})$, given current state $\mathbf{z}$.}
\label{alg:ess}
\begin{algorithmic}[1]
\STATE Draw $\boldsymbol{\nu} \sim \mathcal{N}(\mathbf{0}, \boldsymbol{\Sigma})$.
\STATE Draw an auxiliary variable $u$ and then compute the log-likelihood threshold $\log(y)$:
\begin{gather}
    u\sim \text{Uniform}(0,1)\\
    \log(y) = \log\big(\mathcal{L}(\mathbf{z}-\boldsymbol{\mu})\big) + \log(u)
\end{gather}
\STATE Define the slicing bracket:
\begin{gather}
    \theta \sim \text{Uniform}(0,2\pi)\\
    [\theta_{\text{min}},\theta_{\text{max}}]=[\theta-2\pi,\theta]
\end{gather}
\STATE Compute the proposal state $\mathbf{z'} = (\mathbf{z}-\boldsymbol{\mu}) \cos(\theta) + \boldsymbol{\nu} \sin(\theta)$. \label{alg:draw} 
\IF{$\log\big(\mathcal{L}(\mathbf{z'})\big)>\log(y)$} 
 \STATE Accept the proposal: \textbf{return} $(\mathbf{z'}+\boldsymbol{\mu})$
\ELSE \STATE \textbf{if} $\theta<0$ \textbf{then}  $\theta_{\text{min}}=0$ \textbf{else} $\theta_{\text{max}}=\theta$.
\STATE Draw a new proposal and \textbf{GOTO} \ref{alg:draw}:
\begin{equation}
    \theta \sim \text{Uniform}(\theta_{\text{min}},\theta_{\text{max}})
\end{equation}
\ENDIF
\end{algorithmic}
\end{algorithm}

\section{Summary of proposed MCMC sampler}

Our MCMC sampler is summarized in Alg. (\ref{alg:gibbs}).

\begin{algorithm}[H]
\caption{MCMC sampler for GPIRT}
\label{alg:gibbs}
\begin{algorithmic}[1]
\STATE Initialize the latent trait vectors $\mathbf{x}^{(0)}_{i}$s, slope and intercept parameters $\{\boldsymbol{\beta}^{(0)}_{jt}\}$ and threshold parameters $\{b^{(0)}_c\}$s from their respective priors.
\STATE Initialize latent item function values $\mathbf{f}^{(0)}_{jt}$ by Eq. (\ref{initialize:f}).
\FOR{\text{sampling iteration} $k=0,\dots,K-1$}
    \FOR{\text{item-time} $(j,t)\in \{1,\dots,m\} \times \{1,\dots,T\}$}
    \STATE Sample function values $\mathbf{f}^{(k+1)}_{jt}$ using \acro{ESS} with mean, covariance and likelihood in Eq. (\ref{ess:f}).
\STATE Sample auxiliary latent function values $\mathbf{f^*}^{(k+1)}_{jt}$ over the dense grid $\mathbf{x}^{*}$ using GP posterior rule.
    \ENDFOR
 \FOR{\text{respondent} $i = 1,\dots,n $}
\STATE Sample latent trait vectors $\mathbf{x}^{(k+1)}_{i}$ using \acro{ESS} with mean, covariance and likelihood in Eq. (\ref{ess:x}).
\ENDFOR
\FOR{\text{item-time} $(j,t)\in \{1,\dots,m\} \times \{1,\dots,T\}$}
\STATE Update latent function values $\mathbf{f}^{(k+1)}_{jt}$ to those $\mathbf{f^*}^{(k+1)}_{jt}$ defined on new $\mathbf{x}^{(k+1)}_{t}$.
\STATE Sample slope and intercept parameters $\{\boldsymbol{\beta}^{(0)}_{jt}\}$ using \acro{ESS}.
\ENDFOR
\STATE Sample threshold parameters $\{b^{(0)}_c\}$s.
\ENDFOR 
\STATE \textbf{return} $\{\mathbf{x}^{(k)}_{i}\}_{k=1}^{K},\{\boldsymbol{\beta}^{(k)}_{jt}\}_{k=1}^{K},\{b_{c}^{(k)}\}_{k=1}^{K}, \{\mathbf{f^*}^{(k)}_{jt}\}_{k=1}^{K}$.
\end{algorithmic}
\end{algorithm}

 Table \ref{table:convergence} presents the averaged convergence diagnostics across all three experiments. Our analysis indicates favorable convergence, with all R-hats consistently below $1.1$ and effective sample sizes exceeding $100$. 

\begin{table}[H]
\centering
\caption{{\small Averaged convergence diagnostics across all three experiments. Our analysis indicates favorable convergence, with all R-hats consistently below $1.1$ and effective sample sizes exceeding $100$.}}
\label{table:convergence}
\begin{tabular}{ccccc}
\toprule
Averaged diagnostics   & R-hat of $\mathbf{x}$s  & \acro{ESS} of $\mathbf{x}$s & R-hat of $\mathbf{f}$s   &  \acro{ESS} of $\mathbf{f}$s  \\
\midrule
\acro{simulation} & 1.041   & 160.4  & 1.050 & 185.2 \\
\acro{TAPS}       & 1.057   & 143.2  & 1.086 & 142.2   \\
\acro{Congress}   & 1.012   & 196.7   & 1.010 & 209.0  \\ 
\bottomrule
\end{tabular}
\end{table}

\section{Runtime comparison in \acro{TAPS}}

Although runtime is typically not a primary concern in \acro{IRT} analysis as the collection of data typically requires several months, even years of effort, we document model runtime in \acro{TAPS} application here. In general, all models finished within a reasonable range from a few hours to a couple days. Table \ref{table:taps_runtime}  shows the average runtime of \acro{GD-GPIRT} and baselines on the \acro{TAPS} data. \acro{GD-GPIRT} does require more time to train due to the \acro{MCMC} sampler, but the runtime is still yet quite manageable. 

\begin{table}[H]
    \centering
    \caption{{\small Average runtime of \acro{IPGP} and its competitors in \acro{TAPS} study.}}
    \label{table:taps_runtime}
    \begin{tabular}{lr}
\toprule
\acro{Model} & \acro{Avg Runtime (sec)} \\
\midrule
\acro{GGMU} & 616 \\
\acro{GRM} & 545 \\
\acro{GPCM} & 337 \\
\acro{SRM} & 431 \\
\acro{DSEM} & 333 \\
\midrule
\acro{GPDM} & 200051 \\
\acro{GPIRT} & 222856 \\
\acro{L-GPIRT} & 189977 \\
\acro{DO-IRT} & 39763 \\
\acro{GD-GPIRT} & 343238 \\
\bottomrule
\end{tabular}
\end{table}

\section{Additional application}

We provide an additional application for \acro{GD-GPIRT} in the area of international relation. Respect to human rights is one crucial ingredient of democratic governance \citep{donnelly1999human, beetham1999democracy}. Measurement of governmental human rights practices is usually based on count data of repression activities. One widely recognized dataset is the Cingranelli and Richards Human Rights (CIRI) Dataset which consists of violations of four individual physical integrity variables: political imprisonment, torture, extrajudical killing and disappearances \citep{CIRI2002measure, schnakenberg_fariss_2014}. Each variable is ordinal of scale 3, where high means the right is not violated (0 occurrences), median means the right the right is violated occasionally (less than 50 occurrences) and low means the right is violated frequently (50 or more occurrences).

\begin{figure}[H]
    \centering
     \includegraphics[width=1.0\textwidth]{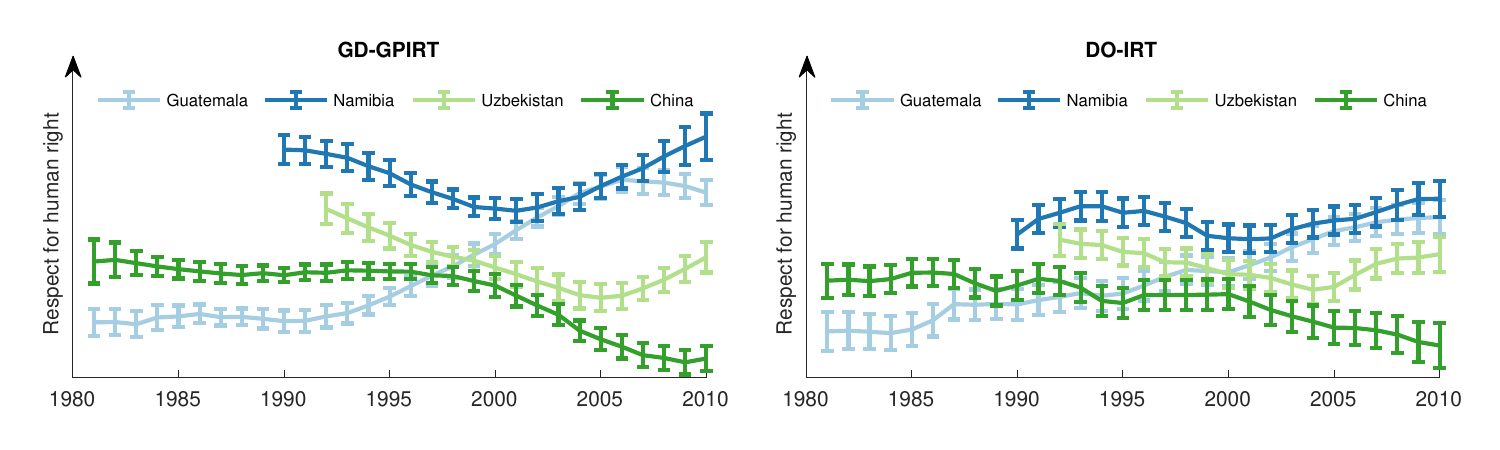}
     \caption{\small{Dynamic human right trends of China, Guatemala, Namibia and Uzbekistan from 1980 to 2010.}}
    \label{fig:CIRIdynamic}
\end{figure}
\begin{figure}[H]

\begin{subfigure}[b]{0.22\textwidth}
              \includegraphics[width=\linewidth]{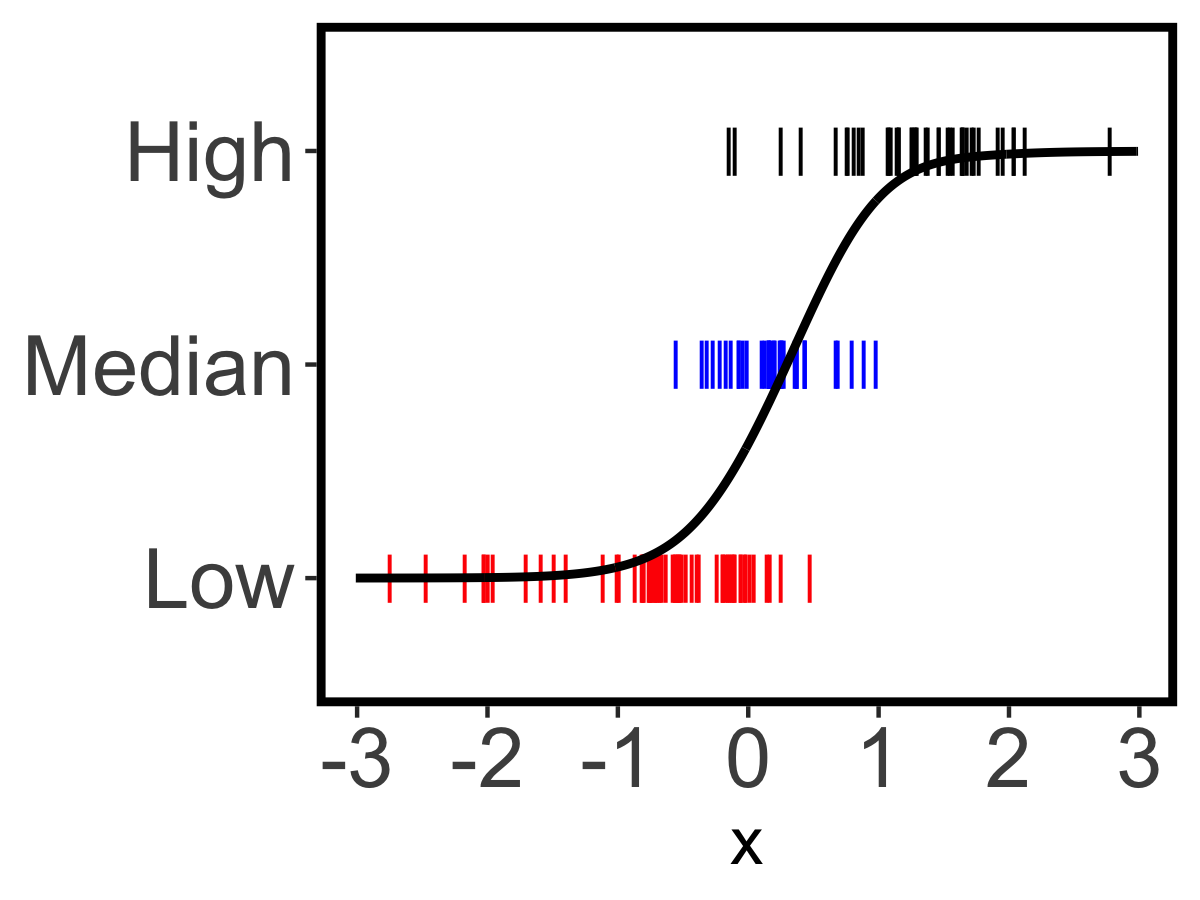}
                \caption{{\footnotesize Imprisonment}}
                \label{fig:CIRIirf1}
        \end{subfigure}\hfill
        \begin{subfigure}[b]{0.22\textwidth}
 \includegraphics[width=\linewidth]{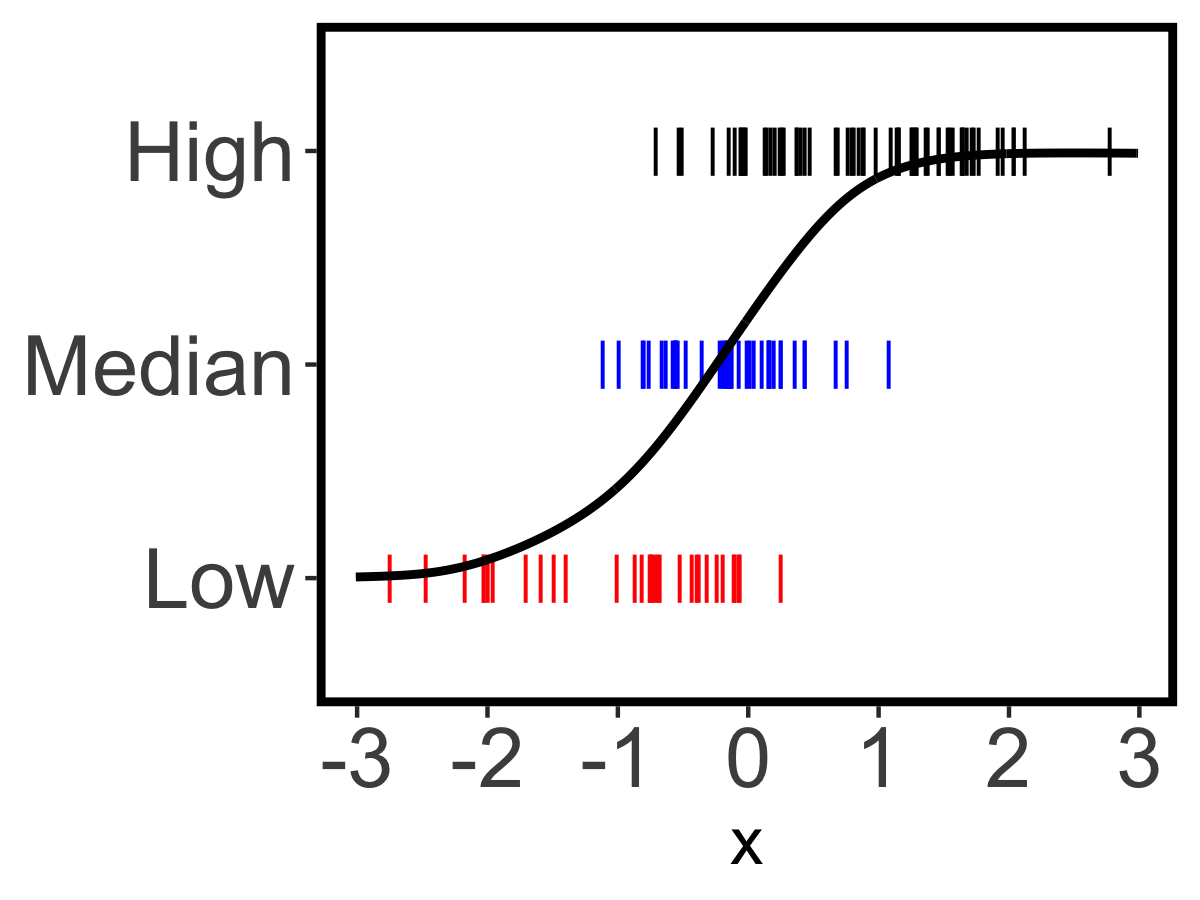}
                \caption{{\footnotesize Torture}}
                \label{fig:CIRIirf2}
        \end{subfigure}\hfill
        \begin{subfigure}[b]{0.22\textwidth}
             \includegraphics[width=\linewidth]{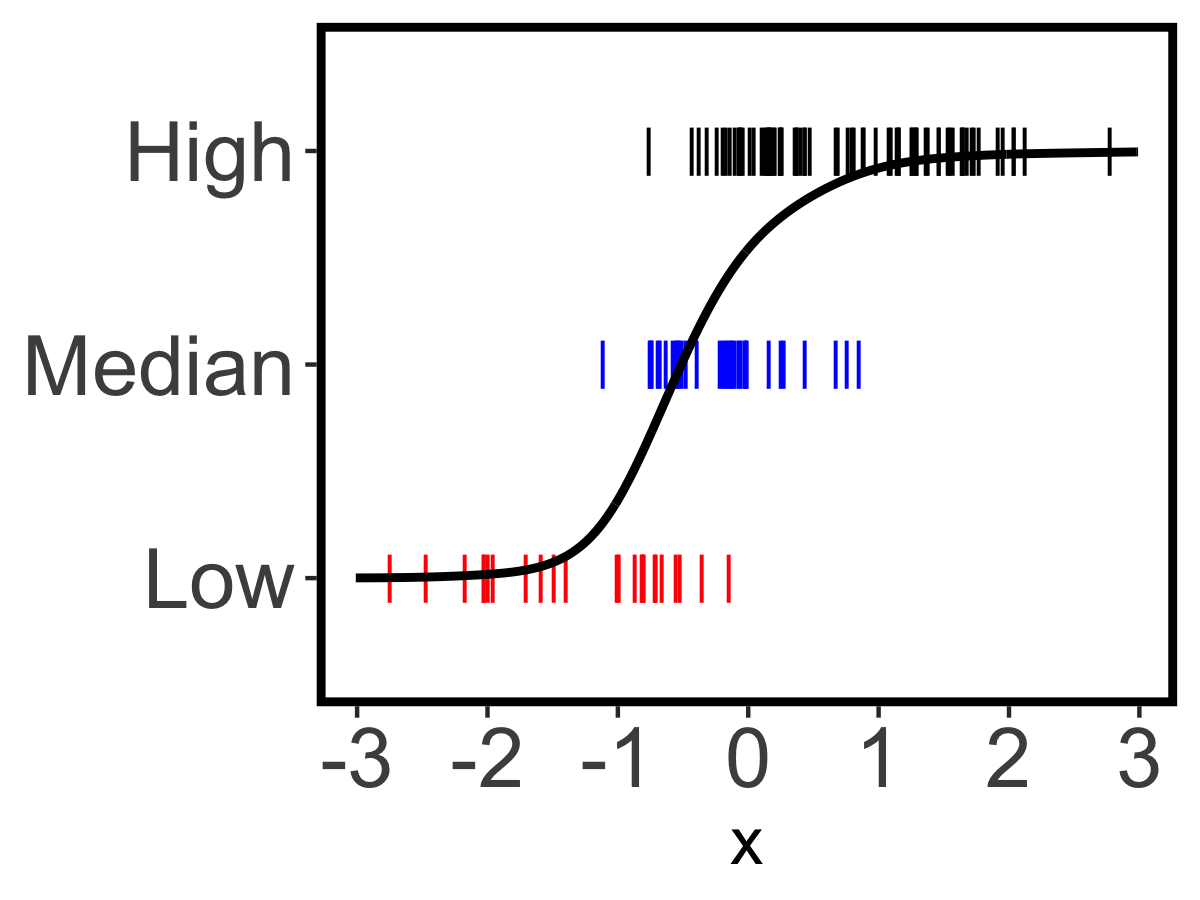}
                \caption{{\footnotesize Extrajudical killing}}
                \label{fig:CIRIirf3}
        \end{subfigure}\hfill
        \begin{subfigure}[b]{0.22\textwidth}
             \includegraphics[width=\linewidth]{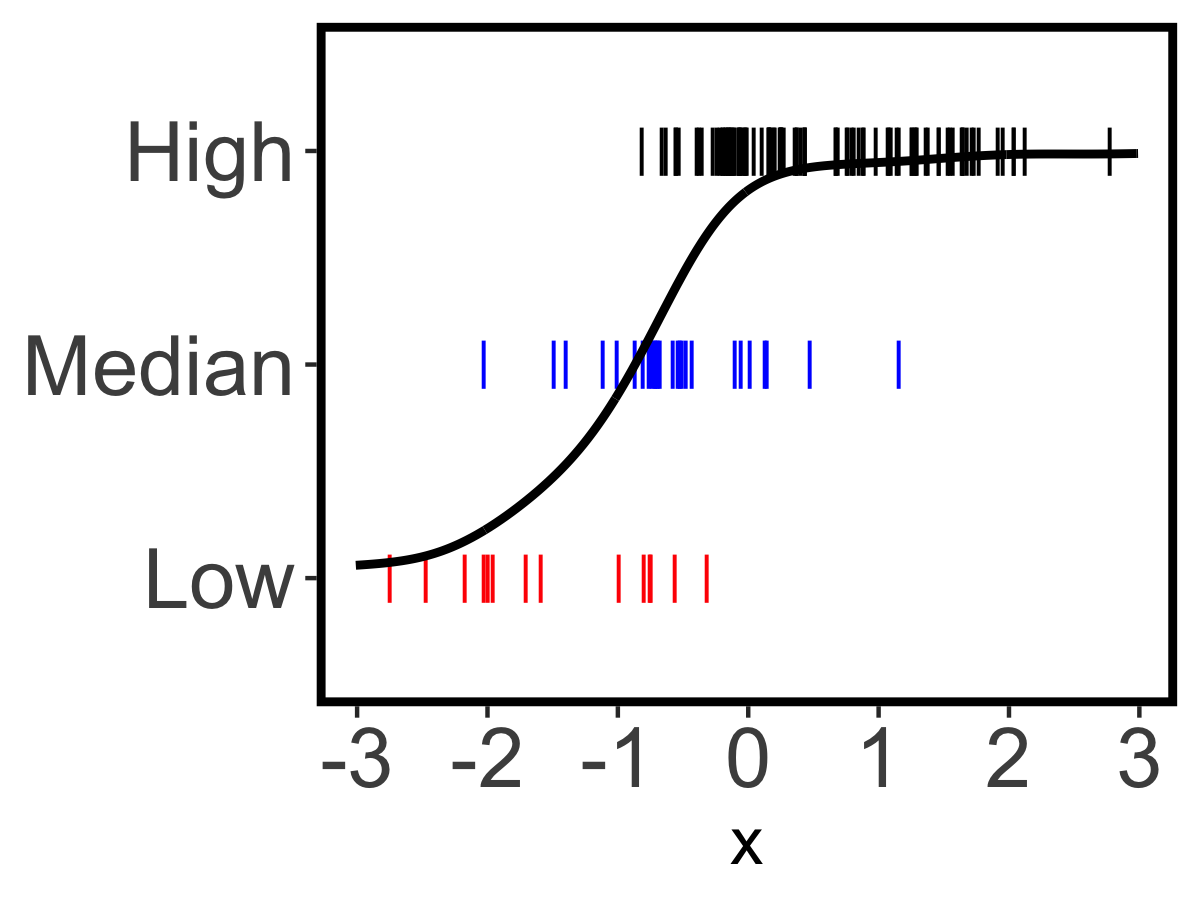}
                \caption{{\footnotesize Disappearance}}
                \label{fig:CIRIirf4}
        \end{subfigure}
        \caption{{\small Estimated \acro{IRF}s for four individual physical integrity variables (political imprisonment, torture, extrajudical killing and disappearances) in 1991 in the CIRI dataset. Expected level of human right indicators is plotted against latent human right score $\theta$. Low, median and high levels of observed human right indicators are displayed as black, blue and red vertical dashes. For all indicators, the estimated human rights levels are roughly monotonic increasing w.r.t to human rights scores as expected.}}
        \label{fig:CIRIirf}
\end{figure}

Estimation of latent human rights scores from repression activity data has been previously studied using standard ordinal \acro{IRT} model \citep{treier2008democracy}. Specifically, \citet{schnakenberg_fariss_2014} extended \acro{D-IRT} to ordinal responses (\acro{DO-IRT}) and augmented the \acro{AR} model with another hierarchical prior on the noise variance. 
We apply \acro{GD-GPIRT} to CIRI data to estimate dynamic latent traits at all time periods simultaneously. Again the length scale of the GP prior on the dynamic latent trends is fixed to the maximum-a-posterior (\acro{MAP}) estimation from a simple GP regression model using \acro{DO-IRT} scores across all countries (7 years). Estimated human right latent traits of \acro{GD-GPIRT} is convergently valid compared to \acro{DO-IRT}, with averaged correlation of $97.4\%\pm 0.6\%$. Figure \ref{fig:CIRIirf} shows estimated \acro{IRF}s for four individual physical integrity variables (political imprisonment, torture, extrajudical killing and disappearances) in 1991 in the CIRI dataset. Expected level of human right indicators is plotted against latent human right score $x$. Low, median and high levels of observed human right indicators are displayed as black, blue and red vertical dashes. For all indicators, the estimated human rights levels are roughly monotonic increasing w.r.t to human rights scores as expected. 

Figure \ref{fig:CIRIdynamic} shows the dynamic human right trends of China, Guatemala, Namibia and Uzbekistan from 1980 to 2010. We also report comparison of model fits between \acro{GD-GPIRT} and \acro{DO-IRT} on CIRI data in Table \ref{table:CIRI}. Since in this application \acro{IRF}s are very much linear, model fits of \acro{GD-GPIRT} only outperforms those of \acro{DO-IRT} slightly.

\begin{table}[H]
\caption{{\small Comparison of model fits between \acro{GD-GPIRT} and \acro{DO-IRT} on CIRI.}}
\label{table:CIRI}
\centering
\begin{tabular}{lrrr}
\toprule
 & \bftab $\mathcal{L}/N$ & \bftab \acro{Acc} & \bftab \acro{AUC} \\
 \midrule
\acro{GD-GPIRT} & \text{\bftab $-$0.540}  & \text{\bftab 0.760} & \text{\bftab 0.793} \\
\acro{DO-IRT} & $-$0.593  & 0.739 & 0.776 \\
\bottomrule
\end{tabular}
\end{table}


\end{document}